Phase-stabilized 100 mW frequency comb near 10 μm

Kana Iwakuni[1], Gil Porat[1], Thinh Q. Bui[1], Bryce J. Bjork[1†], Stephen B. Schoun[1], Oliver H. Heckl[1*], Martin E. Fermann[2], Jun Ye[1]

[1] JILA, National Institute of Standards and Technology and University of Colorado, Department of Physics, University of Colorado, Boulder, CO 80309, USA

[2] IMRA America, Inc., 1044 Woodridge Ave., Ann Arbor, Michigan 48105, USA

†Present address: Honeywell International, 303 Technology Court, Broomfield, CO 80021, USA
*Present address: Christian Doppler Laboratory for Mid-IR Spectroscopy and Semiconductor Optics, Faculty of Physics, University of Vienna, Boltzmanngasse 5, 1090 Wien

e-mail kana.iwakuni@jila.colorado.edu

**Abstract:** Long-wavelength mid-infrared frequency combs with high power and flexible tunability are highly desired for molecular spectroscopy, including investigation of large molecules such as $C_{60}$. We present a high power, phase-stabilized frequency comb near 10 μm, generated by a synchronously-pumped, singly-resonant optical parametric oscillator (OPO) based on $AgGaSe_2$. The OPO can be continuously tuned from 8.4 to 9.5 μm, with a maximum average idler power of 100 mW at the center wavelength of 8.5 μm. Both the repetition rate ($f_{rep}$) and the carrier-envelop offset frequency ($f_{ceo}$) of the idler wave are phase-locked to microwave signals referenced to a Cs clock. We describe the detailed design and construction of the frequency comb, and discuss potential applications for precise and sensitive direct frequency comb spectroscopy.

1. Introduction

High power frequency comb sources at infrared wavelengths have become increasingly desirable for a wide range of applications in spectroscopy. While originally developed for metrology, comb lasers found immediate applications in high-resolution spectroscopy [1, 2]. In direct frequency comb spectroscopy, frequency combs are used as light sources, enabling efficient and multiplexed, high-resolution spectroscopy over a broad spectral bandwidth [3]. Direct frequency comb spectroscopy based on a variety of broad-band detection schemes including Fourier-transform spectrometry (FTIR) [4], grating spectrography [5], virtually-imaged phased array (VIPA) spectrometry [6], and dual-comb spectrometer [7] have been developed mainly for applications of trace gas sensing [8, 9]. Recently, the versatility of direct comb spectroscopy was extended to probe large, complex molecules cooled by buffer gas [10] and chemical reaction kinetics involving transient intermediates such as DOCO molecules produced from the OD + CO reaction [11, 12]. These studies were carried out in the mid-infrared (MIR) wavelength (3 – 5 μm), and the next advance for comb-based high-resolution molecular spectroscopy lies at longer mid-IR wavelengths of 5 – 10 μm, where the fundamental molecular vibration bands display even stronger intensities. In the case of large molecules, probing at longer wavelengths provides an additional advantage of reducing the impact of intermolecular vibrational redistribution, which causes spectral congestion and dilution of absorption intensity [13]. For detection of transient molecules like DOCO, strong band intensity is also necessary since the probing time must be very short, on the order of microseconds, to follow kinetics.

Recently, frequency comb lasers in the MIR wavelength region have been reported using a variety of architectures: difference-frequency generation (DFG), optical parametric oscillators (OPO), quantum cascade lasers (QCL), microresonator, and supercontinuum using waveguide dispersion.

DFG-based mid-IR frequency comb sources (based on periodically-polled lithium niobate, PPLN) are known for their relative simplicity (single-pass configuration, passive cancelation of the $f_{ceo}$) at the expense of relatively low power (100 μW–20 mW at 2.6 – 5.2 μm) [14]. More recently, a higher power (~ 240 mW) DFG comb has been reported at 2.7 – 4.2 μm based on Raman-induced soliton self-frequency shift in a nonlinear fiber [15]. Longer mid-IR wavelengths which are comparable to those in the present work have been achieved with a GaSe crystal at 7.5 – 12.5 μm (~ 15 μW) [16] and 8 – 14 μm (~ 4 mW) [17], with a $AgGaS_2$ crystal at 7.5 – 11.6 μm (~ 1.55 mW) [18], and with an orientation-patterned (OP) GaP at 6 – 11 μm (~ 60 mW) [19].

In contrast to DFG-based sources, OPO-based frequency combs exhibit higher power while maintaining a wide wavelength tuning range. Mid-IR sources below 6 μm have been reported using MgO-PPLN at 2.8 – 4.8 μm (~ 1.5 W) [20] and at 2.2 – 3.7 μm (~ 33 mW) [21], using OP-GaAs at 2.6 – 7.5 μm (~ 73 mW) [22] and at 5.2 – 6.2 μm (~ 10 mW) [23], using $AgGaSe_2$ at 4.8 – 6.0 μm (~ 17.5 mW) [24], and using OP-GaP at 2.3 – 4.8 μm (~ 30 mW) [25]. There have been several reported implementations of longer wavelength infrared OPOs at 5 – 12 μm (~ 15 mW at 8.5 μm) [26], and 2.85 – 8.4 μm [27], but the idler powers in these studies drop off significantly at wavelengths longer than 8 μm. Finally, the widest wavelength tuning (1.33 – 20 μm, ~ 40 mW at 9 μm) has been recently reported by combining OPO and DFG techniques [28].

Additional infrared frequency comb sources include QCL, microresonators and supercontinuum architectures. A QCL-based frequency comb has a high repetition rate (around 10 GHz) and high power (~ 3 mW per comb mode) for both 7 μm [29] and 8 μm [30] center wavelengths, though it is difficult to tune its wavelength continuously. On the other hand, microresonator [31, 32, 33] and supercontinuum frequency combs [34, 35, 36] have been limited to wavelengths no longer than 3 μm. Access to longer infrared wavelengths are mainly hindered by the difficulty of finding suitable materials.

As described above, ideal light sources for molecular spectroscopy in the long infrared wavelengths, i.e., frequency comb sources providing a combination of high power, frequency stability, and broad wavelength tunability, are highly desired for molecular spectroscopy in the long infrared wavelengths. In this work, we have developed a singly resonant $AgGaSe_2$-based OPO synchronously pumped with a Tm fiber comb. We achieved more than 100 mW average power near the 10 μm region, for the first time to the best of our knowledge. Also, we experimentally demonstrate phase-stabilization of the frequency comb. This long wavelength infrared laser will enable the study of fundamental vibration bands of large molecules and molecular clusters, which are important for investigating structure and dynamics of complex molecules.

## 2. Apparatus and results

### 2.1 Optical parametric oscillator

Figure 1 depicts the experimental setup. The pump laser is a mode-locked Tm:fiber laser at 1.95 μm, with a spectral bandwidth of 50 nm, repetition rate of 110 MHz, and a maximum output power of 2.5 W. This Tm-fiber laser synchronously pumps a 5-mirror linear OPO cavity, where all mirrors have high reflectivity (HR, >99.5%) at the signal wavelength for singly-resonant operation. Mirror M4 out-couples the generated idler wave with a transmittance of about 92%, while M1 functions as an input mirror with about 99% transmittance for the pump wave. M5 is attached to a fast piezo (PZT) actuator for $f_{ceo}$ control [37].

The OPO is based on a $AgGaSe_2$ nonlinear optical crystal with an aperture of 5×5 mm and length of 3.6 mm. The crystal is centered at the focal plane of the concave mirrors, M2 and M3, and mounted on a rotation stage for angle tuning. Both faces of the crystal are anti-reflection (AR) coated for the pump, signal, and idler waves. The $AgGaSe_2$ crystal has a relatively large second order nonlinear optical coefficient of 32 pm/V, and it is transparent at 0.76 – 17 μm (absorption coefficient < 1 $cm^{-1}$) [38]. These characteristics make $AgGaSe_2$ highly suitable for mid-infrared generation using a 1.95 μm pump wave while avoiding large nonlinear absorption.

This is in contrast to GaAs, where the idler power was significantly limited due to nonlinear absorption of the pump and signal [23]. The AgGaSe$_2$ crystal is cut at an angle of 57° to achieve type II phase matching with pump wave (1.95 μm, extraordinary wave), signal wave (2.53 μm, ordinary wave), and idler wave (8.5 um, extraordinary wave).

The generated idler power is determined by a compromise between the nonlinear interaction strength, effective crystal length, and thermal lensing, under the limitation imposed by the crystal's optical damage threshold. The nonlinear interaction strength increases with peak pump intensity, which in turn is greater for shorter pulse width and smaller beam size. However, as the pulse gets shorter and the spot size smaller, other detrimental effects limit the OPO performance. First, a shorter pulse width reduces the effective interaction length due to temporal walk-off [39] in the crystal. Second, a smaller spot size exacerbates thermal gradients across the crystal, resulting in a thermal lensing effect. The latter effect is more severe in AgGaSe$_2$ relative to most common nonlinear crystals as it is difficult to cool AgGaSe$_2$ due to its low thermal conductivity (1 W/(m・K)) and large size. The large crystal size is a result of its softness (Mohs hardness of 3), making it challenging for further machining process to yield small pieces. The combination of a large size and low thermal conductivity makes it challenging to remove optically deposited heat from the crystal. In addition to these trade-offs, the peak intensity is limited by the crystal's damage threshold, which we observed to be about 3 GW/cm$^2$ for ~ 100 fs pulses. In the following, we will discuss our comprehensive experimental investigations of these competing effects and present a final design that yields an optimal idler power of more than 100 mW.

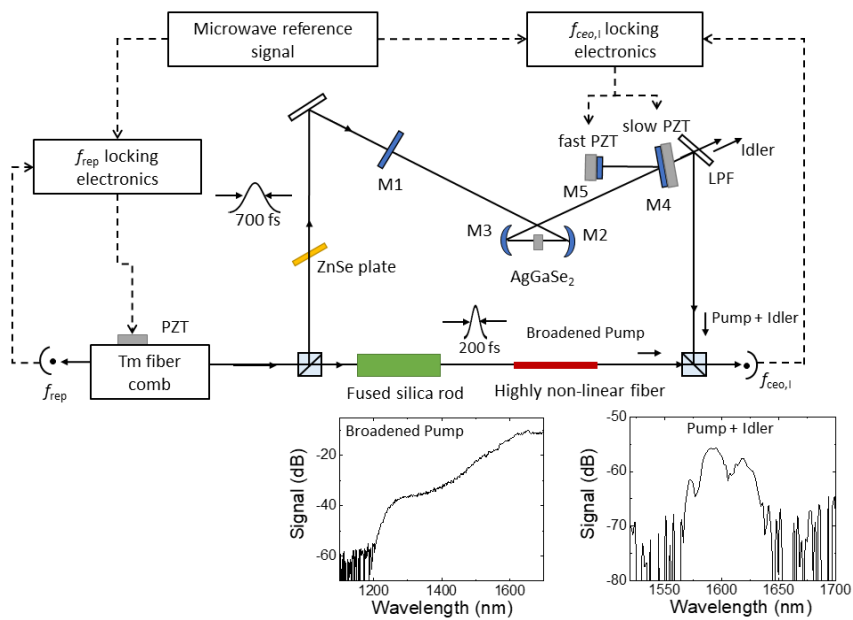

Fig. 1 Schematic of the OPO. Output from a 110 MHz Tm fiber comb is mode-matched into linear cavity consisting of the M1, M2, M3, M4 and M5 mirrors. M2 and M3 are concave mirrors with a ROC of either 50 mm or 110 mm (see text). The rest are flat mirrors. All mirrors have HR for the signal wave. M1 has high transmittance (HT) for the pump wave and M4 has HT for the idler wave. A part of Tm fiber comb light enters highly non-linear fiber after pulse compression by the fused silica rod, and it is spectrally broadened. P + I and spectrally broadened P interfere to obtain $f_{ceo,I}$. LPF stands for long pass filter.

**2.2 Idler power optimization**

The line plot in Fig. 2(a) shows the calculated average pump power at the oscillation threshold of the OPO as a function of the pump pulse width [40]. This calculation makes the assumption that both the pump and signal waves have 30 μm ($1/e^2$ intensity radius) beam sizes and the same pulse width. The estimated round-trip loss of the OPO cavity is 14.7%, calculated from mirror reflectivities and crystal linear absorption. The calculation shows that there is the trade-off between pump pulse width and effective crystal length (shown in the inset of Fig. 2(a)). The experimental OPO threshold under three different pulse widths are measured and plotted in Fig. 2(a) with the same plot markers of Fig. 2(b). The measurements basically follow the theory prediction and show that the lowest threshold is achieved at 700 fs.

Figure 2(b) shows the measured average idler power as a function of input pump power at three different pump pulse widths. At the crystal, the pump and signal beam radii are set to ~ 30 μm by using 100 mm radius of curvature (ROC) concave mirrors (M2 and M3). With this beam size, the maximum input pump power is limited to 1.2 W due to the low damage threshold of the crystal AR coating. The 200 fs pulse width, obtained by pulse compression with a fused silica rod, is the shortest studied here. We also stretch the pulse by inserting ZnSe plates or a fused silica rod to the pump beam's path (see Fig. 1). When using a 5-mm thickness ZnSe plate, a stretched pulse of 700 fs yields the maximum observed idler power of 105 mW. The photon conversion efficiency is about 60% at the conditions that produces the maximum idler power.

We observe saturation of the output idler power at increasing average pump power. This saturation behavior is evident for all three traces in Fig. 2(b) when the average pump power exceeds 800 mW. A likely explanation is thermal lensing, where the thermal lens focal length is inversely proportional to average intensity. The latter increases for higher average power and smaller beam size. Figure 2(c) shows the average idler power as a function of pump power for two different beam sizes of 20 μm and 30 μm, both using a pulse width of 200 fs. For a beam size of 20 μm (obtained from using concave mirrors with ROC of 50 mm), there is strong saturation of the idler power above 350 mW pump power as shown in Fig. 2(c), which corresponds to the same average intensity for 800 mW at 30 μm beam size. Thermal lensing could be mitigated by cooling of the crystal. However, this strategy is not effective due to the crystal's large thermal mass and low thermal conductivity. These results suggest that larger ROC mirrors could further relax this thermal effect and increase idler power, especially in combination with a more powerful pump laser.

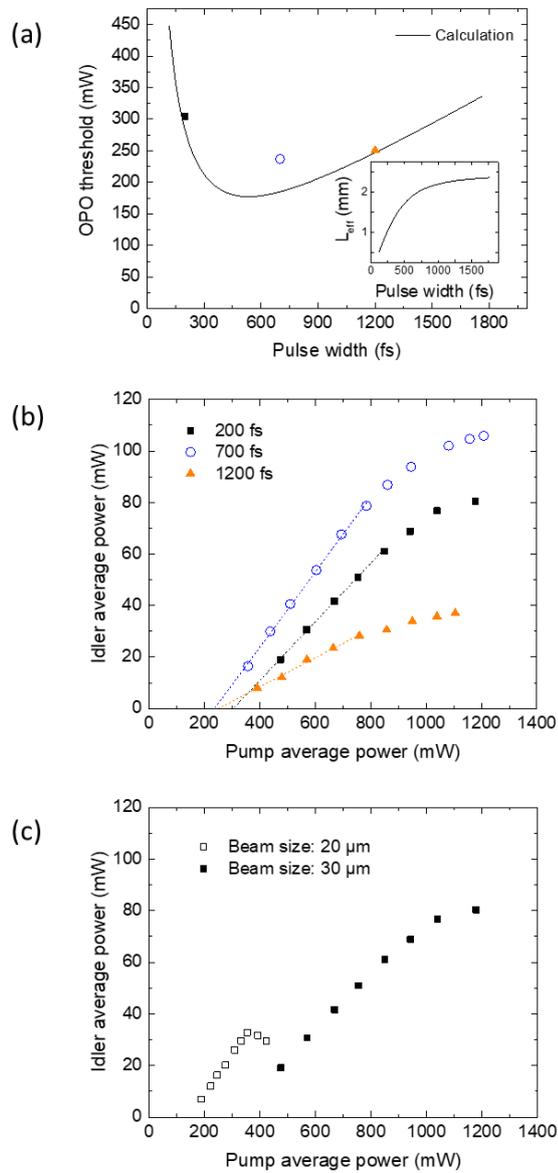

Fig. 2 OPO oscillation threshold and idler power. (a) Calculated average pump power at the oscillation threshold of the OPO as a function of pump pulse width (FWHM), with a beam size

of 30 μm ($1/e^2$ intensity radius, solid line). The data points show the measured threshold for each pulse width shown in Fig. 2(b). Inset shows the effective crystal length ($L_{eff}$) as a function of the pump pulse width (FWHM). $L_{eff}$ is limited by spatial transverse walk-off at long pulse widths. (b) Idler average power as a function of pump average power at different pulse widths. The dashed lines indicate trends before saturation starts. (c) Idler average power as a function of pump average power at different beam sizes ($1/e^2$ intensity radius), with a pulse width of 200 fs (FWHM).

**2.3 Wavelength tuning**

The OPO's wavelength can be tuned by changing the crystal temperature and phase-matching angle. We first studied temperature tuning. The crystal refractive index has a small temperature dependence of $dn/dT = 8 \times 10^{-5}$ $K^{-1}$ [38]. We observed a very limited idler wavelength tuning range of about 1 nm by temperature tuning from 20 °C to 60 °C due to the narrow safe operating temperature range for the crystal's AR coating.

The phase-matching angle provided a significantly larger wavelength tuning capability. Figure 3 shows the output idler spectrum recorded with a Fourier-transform infrared spectrometer [41] for different phase-matching angles from 54.2° - 57.2° (solid lines). This corresponds to center idler wavelengths of 8.4 - 9.5 μm with a spectral bandwidth of 200 nm (FWHM). The average idler power corresponding to each spectrum's center wavelength is overlaid on the same plot depicted by black square points. The strong absorption lines present in the spectra are from the $v_2$ fundamental vibrational band of atmospheric water. Fringes on the spectra are due to etalon effects from optics outside of the OPO cavity. Access to longer idler wavelength is limited by the HR coating of the M1 mirror for the signal, while the shorter idler wavelength is limited by absorption of atmospheric water at the signal wavelength, which increases the loss of the OPO cavity and hence degrades the idler power (Fig. 3b).

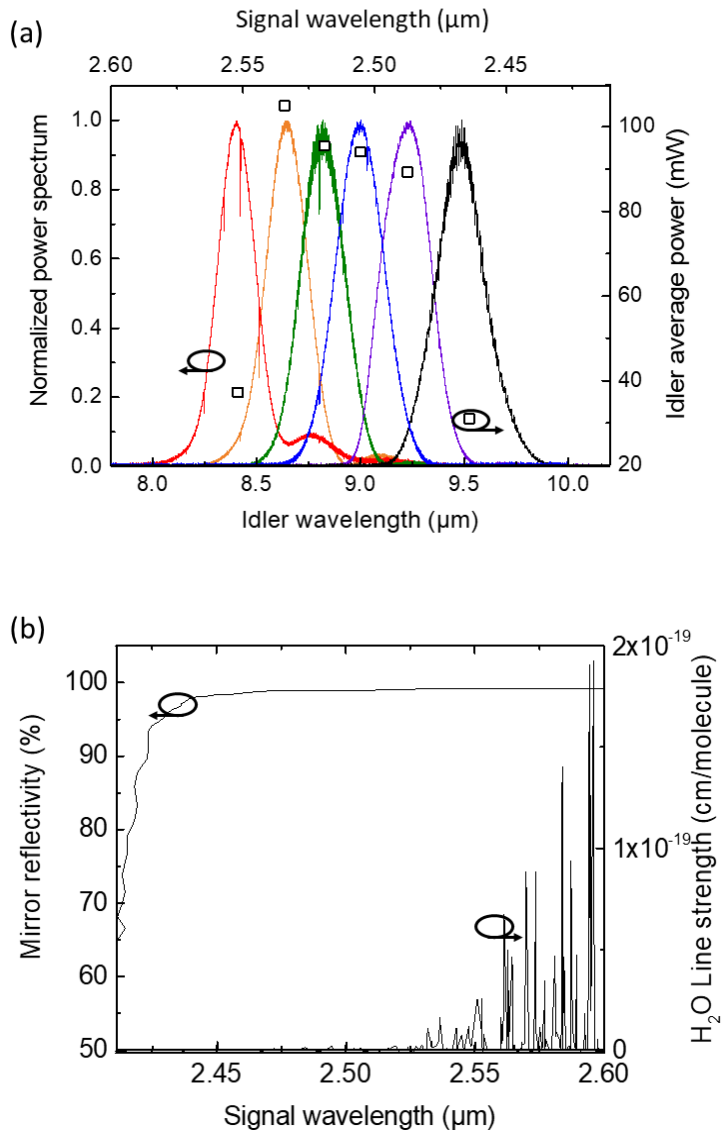

Fig. 3 OPO wavelength tuning. (a) Idler spectrum and corresponding idler average power. (b) The reflectivity of input mirror M1 (left) and the line strength of water rovibrational features (right) as a function of signal wavelength.

## 2.4 $f_{rep}$ and $f_{ceo}$ phase stabilization

The idler frequency comb can be completely stabilized by phase locking $f_{rep}$ and $f_{ceo}$ to a radio frequency (RF) reference signal. To lock $f_{rep}$, a small fraction of the output light from the

Tm comb oscillator is detected by a fast photodiode. The 9th harmonic of $f_{rep}$ (990 MHz) is mixed with a 1 GHz reference frequency from a Cs-stabilized quartz oscillator. The resulting 10 MHz beat signal is measured by a phase detector, where it is mixed with a reference RF signal generated by a DDS slaved to a 10-MHz quartz oscillator stabilized to a Cs clock. The phase error signal from the mixer is low-pass filtered and fed back to a PZT in the Tm comb oscillator to lock $f_{rep}$ (Fig.1).

Stabilization of the carrier envelop offset frequency of the idler wave, $f_{ceo,I}$, is achieved by the heterodyne optical beat of 1) the sum frequency of the pump and idler wave (P + I) generated in the OPO cavity and 2) spectrally broadened pump wave (P) by a highly nonlinear fiber [20]. The P + I wave at 1.6 μm, along with many other parasitic waves, are generated in the crystal by non-phase-matched nonlinear processes. The bottom insets of Fig. 1 show the broadened pump spectrum from a 24-cm long, 4-μm core diameter nonlinear fiber and the P + I spectrum from the OPO. To broaden the pump spectrum efficiently, the pulse width is narrowed to 200 fs with a 30-cm fused silica rod. The pump spectrum broadens to 1.6 μm at 300 mW input power into the nonlinear fiber.

The $f_{ceo,I}$ beat signal is detected with a single photodetector with an input power of approximately 50 μW. The signal-to-noise ratio of the beat signal is limited by the amplitude noise from the spectrally broadened pump wave [42], which was suppressed by 5 dB using balanced detection. We achieved a final signal-to-noise ratio of 35 dB at 100 kHz resolution bandwidth. Figure 4(a) shows the observed free-running $f_{ceo,I}$ beat signal. $f_{ceo,I}$ is phase-stabilized to the same Wenzel quartz oscillator used for $f_{rep}$ locking. The error signal from the phase detector is fed back to both a fast PZT (M5) and slow PZT (M4) in the OPO cavity (Fig. 1). Figure 4(b) shows the in-loop beat signal of $f_{ceo,I}$ with $f_{rep}$ phase-locked. The servo bandwidth is about 40 kHz based on Fourier analysis of the error signal. The coherent peak is about 30 dB with a spectral width limited by the instrumental resolution bandwidth of 1 kHz. The estimated linewidth of the idler wave based on the linewidth of the quartz oscillator (~ 50 μHz) is a few kHz [43].

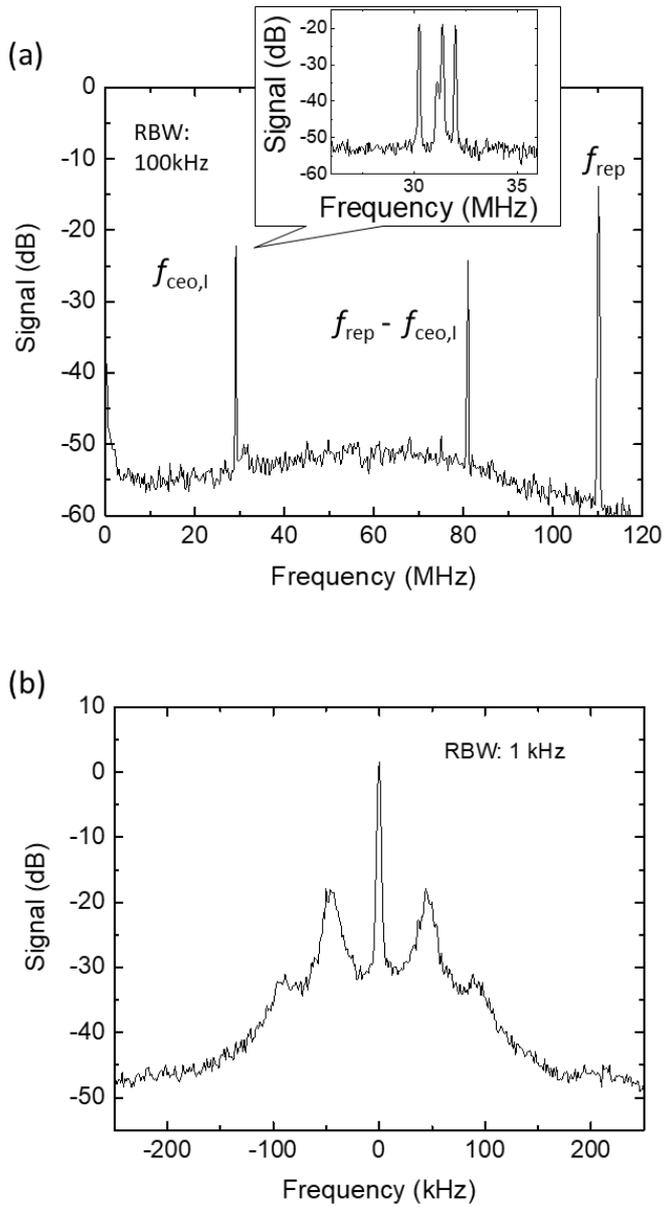

Fig. 4 $f_{ceo}$ phase-locking. (a) Free-running $f_{ceo,I}$ beat signal at a resolution band width (RBW) of 100 kHz and sweep time of 5 ms. Inset shows a magnified view of the free-running beat signal at a RBW of 100 kHz and sweep time of 30 ms. (b) Phase-locked in-loop beat signal at a RBW of 1 kHz. The linewidth is limited by the RBW and the servo bump is at 40 kHz.

## 3. Conclusions

We have developed a phase-stabilized frequency comb generated from an OPO in the wavelength range of 8.4-9.5 μm. We obtain a maximal idler power of more than 100 mW at maximum (10 μW per comb mode) by optimizing the pulse width and beam size, which is the highest reported power in this wavelength region. In addition, we find that the thermal lensing effect becomes significant at an average intensity of about 28 kW/cm$^2$. Currently, the idler power is limited by the thermal effects and damage threshold of the crystal. These limits can be further relaxed by using mirrors with a larger ROC. In wavelength tuning, we expect that a 10 μm idler wave can be obtained with the current crystal by only extending the mirror HR coating up to 2.4 μm. Shorter idler wavelengths (< 8.4 μm) can be obtained by properly purging the OPO cavity with dry air. Finally, we demonstrate that the idler's $f_{rep}$ and $f_{ceo}$ are phase-locked to a stable microwave reference. In the future, we plan to couple the idler into an enhancement cavity and perform multiplexed readout of the output spectrum with highly-dispersive optics, such as a VIPA. These combined techniques will be a significant upgrade for our goals of performing high-resolution spectroscopy of large molecules such as $C_{60}$ and time-resolved detection of transient molecules.

## Acknowledgements

This research was funded by AFOSR, DARPA SCOUT, NIST, and NSF-JILA PFC. K.I. is supported by the JSPS Overseas Research Fellowships. T.Q.B. is supported through an NRC Postdoctoral Fellowship. We would like to thank P. Bryan Changala for helpful discussions.